# Sliding van der Waals Polytypes

Maayan Vizner Stern[1#], Simon Salleh Atri[1#], Moshe Ben Shalom[1*]

[1]School of Physics and Astronomy, Tel Aviv University, Israel

Compared to electronic phase transitions, structural phase transitions of crystals are challenging to control due to the energy cost of breaking dense solid bonds. Recently, however, electric field switching of stacking configuration between honeycomb layers, held together by relatively weak van der Waals (vdW) attractions, was demonstrated. In response to the external fields, the layers slide between commensurate meta-stable configurations with discrete symmetries and distinct lattice orientations. These 2D vdW polytypes host diverse electronic orders such as ferroelectricity and magnetism, providing multiferroic switching via lubricant sliding of incommensurate boundary strips.

Ahead, we address recent observations in honeycomb polytypes and identify remaining challenges for extending this conceptual "SlideTronics" mechanism into rapid, local, and practical multiferroic devices. The stacking energies, symmetries, and orbital overlaps that underlie the band structures and internal charge distributions are discussed, along with poly-properties like interfacial-ferroelectricity, ladder-like cumulative polarization, superconductivity, and orbital magnetic orders. Distinct from conventional 3D multiferroic crystals, the 2D vdW assembly and the sliding switching mechanism open poly-opportunities for novel device concepts.

**Introduction:**

The Field Effect Transistor (FET) [1] is acknowledged as a core concept enabling ongoing progress in science and technology across many fields. A local electric field is applied to perform a computation, populate quantum Hall states [2], and much more by changing the density of a two-dimensional (2D) electron gas. Recently, it was found that a field effect device can further switch the atomic positions of a van der Waals (vdW) interface between layered crystals [3–7]. Such electric field control of structural transitions is highly appealing [8,9] because atomic positions and crystal symmetries impact collective properties beyond electrical conductivity. Indeed, research efforts devoted to different vdW stacking configurations over the past few years revealed numerous structurally coupled responses. Examples include cumulative ferroelectricity [10,11], spin order [12–15], orbital magnetism [16,17], unconventional superconductivity [18–20], and fractional quantum anomalous Hall effect [21]. Exploiting the full potential of these switchable 2D vdW polytypes requires a deeper understanding and subsequent control of the structural sliding transitions and their coupling to electronic orders.

Towards these goals, we describe the poly-structural symmetries available in honeycomb vdW polytypes and the underlying electronic states that govern the poly properties [22]. Then, we address elastic relaxation processes that impose poly-structural stabilities and enable the assembly of all metastable periodic configurations in adjacent domains, separated by narrow dislocation strips (unlike 3D polytypes). The relative stability of each configuration under external electric and stress fields and the mobility of the boundary strips enable unique switching dynamics [23,24]. We point to the main obstacles restricting domain nucleation and strip mobility and to novel concepts to overcome these bottlenecks [25,26]. By combining these recent results, we hope to attract the electronic and elastic-oriented research communities towards poly-opportunities beyond 3D crystals that lie ahead.

**Switching Polymorphs, Polytypes, and 2D vdW Polytypes**

Carbon atoms may condense into a graphite polymorph of honeycomb layers stacked at the tip of a pencil with soft, opaque, semi-metallic characteristics (see Fig. 1a). Alternatively, the carbons may form a Diamond polymorph with stiff, transparent, and insulating properties. Now, imagine a gentle press that switches the tip of your pencil into a diamond. Clearly, it would be remarkable (and profitable), especially if controlled efficiently, reversibly, and rapidly as in electronic devices. Unfortunately, turning graphite into diamond is impractical due to the high pressure and temperature required to activate 3D rearrangements of carbon bonds [27–30], despite having similar condensation energy [31,32].

**Polytypes** are polymorphs with distinct atomic orders along one of the crystal dimensions only [33] and hence are more likely to divide into local domains. For example, SiC polytypes share identical honeycomb planes of silicon and carbon atoms stacked in periodic sequences (as in graphite) [34]. While the stacking sequence can tune the optical response of each SiC polytype [35], growing a particular one on macroscopic scales is challenging due to the sensitivity of 3D growth process to nucleation conditions [36]. Switching 3D polytypes is even more challenging [37,38]. First, a structural domain must nucleate, separated by a 2D boundary dislocation that accommodates energy-expensive intermediate atomic configurations. Then, the boundary wall should move across the system to extend one polytype configuration over another. The reduced atomic density and the loss of 3D binding energy at the wall dislocation restricts a practical electrical switching of structural phases in covalently (or ionic) bonded polytypes at temperatures away from the melting point, even if applied to thin film structures [39,40]. For preexisting dislocations, further switching barriers arise due to local contaminants and disorder, tending to pin the boundary wall motion [41].

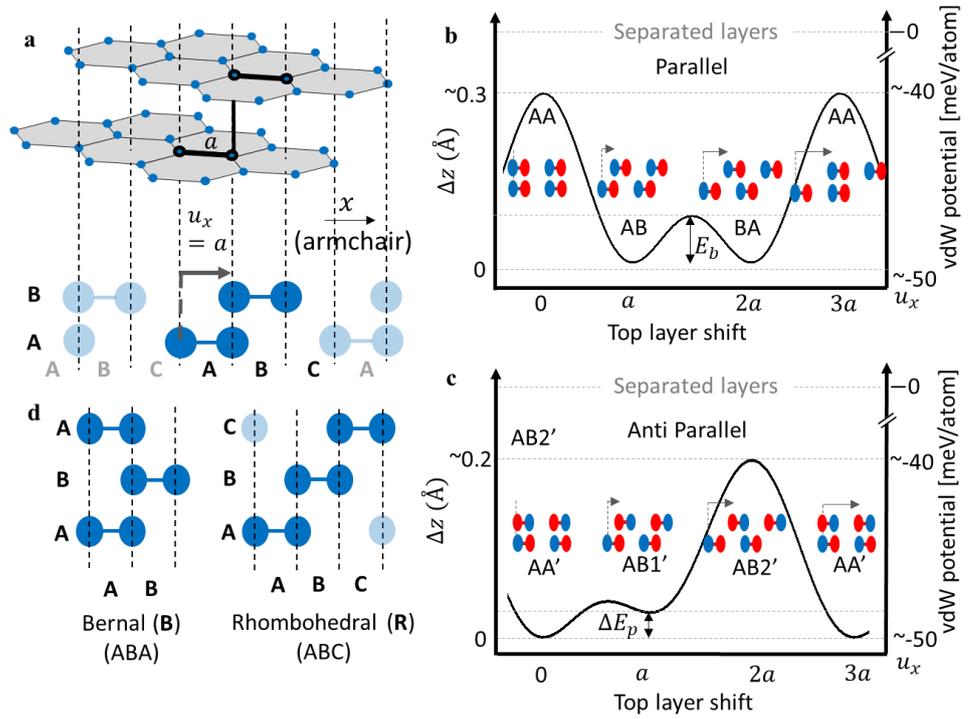

*Figure 1: Stacking potentials* (a) AB polytype illustration showing interlayer shift by $a$, the inter-site distance, along the armchair direction. A unit cells with two atoms per layer is emphasized (b) Typical stacking energies (right axes) and vertical separation variation $\Delta z$ (left) as a function of the planar shift $u_x$ [42–44]. The depth of the potential well $E_b$ is marked by arrows (c) Interfacial stacking potential in anti-parallel binary compounds.[45–48] (d) ABA and ABC polytypes of trilayer graphene.

We refer to **2D vdW polytypes** as a finite number N of layers attached by vdW forces into high symmetry stacking configurations of (meta) stable periodic crystals. Owing to a relatively low stacking energy per atom, small number of atoms at 1D boundary strips, and pristine super-lubricant interfaces[49] free from pining disorder and contaminants (segregated into isolated bubbles) [50,51], the structural configuration of these polytypes can switch using electric field only. We start by describing the poly-structural symmetries of honeycomb graphitic polytypes (carbons only) or binary compound polytypes with distinguished atoms like boron-nitride (BN) and transition metal dichalcogenides (TMD). Then, we discuss their poly-properties arising from the different overlaps of interfacial orbitals, electronic band hybridizations, density of states (DOS), and internal charge redistributions. We emphasize the poly-structural stabilities and in-plane atomic relaxation mechanisms that guarantee the poly-assembly of all adjacent configurations, including inequivalent polarization orientations. Any marginal interfacial strain or twist squeezes into narrow partial dislocation strips, separating structural domains of uniform configurations. We elaborate on the poly-switching dynamics and the macroscopic response governed by the microscopic boundary details, the strip network, and the interaction between strips from adjacent interfaces. Finally, we point to the remaining switching obstacles, such as dislocation nucleation[25] and local strip pining by open bonds at the edge[23] that restrict practical multiferroic opportunities.

### Sliding 2D vdW polytypes

Consider two layers of graphene aligned precisely one above the other, with the carbon atoms of one layer positioned perfectly above the atoms from the other layer. Such a fully eclipsed AA stacking configuration is unstable [42,52,53] and therefore, one of the layers naturally slides by $u_x = a$, the honeycomb inter-atomic distance along the armchair direction (see the rigid spheres illustration Fig. 1a). The eclipsed/hollow carbon environment in this shifted AB polytype configuration force electronic rearrangement between the two inequivalent atomic sites that reduce the interlayer distance Z to improve the packing density [54–56]. In particular, the relatively low Pauly repulsion in the hollow sites (facing empty hexagon center) creates a stacking potential well, essential for the structural stability as shown in Fig. 1b. Along a periodic sliding cycle $0 < u_x < 3a$, the adhesion potential varies by ~ ten meV per atom, while the interlayer distance changes by $\Delta Z \sim 0.2$Å [42–44,57,58] (right and left axes, respectively). Note the structural anisotropy of layered vdW polytypes with a typical distance $a \sim 1$Å between the atoms within each layer versus substantially longer distance between the layers Z ~3Å, and correspondingly intra/interlayer binding energies of ~1000 / 50 meV per atom (see the zero-adhesion potential for "separate layers")[59]. Also, note the minute atomic displacement out of the plane within each layer < 0.01Å due to high planar stiffness. Additional sliding by inter-atomic distance to $u_x = 2a$ results in an identical BA polytype that differs by indistinguishable rotation of 60 ($+120n, n \in \mathbb{N}$) degrees in the plane.

Next, consider different atoms in each layer, such as the monolayer binary-compound hexagonal boron nitride (*h*-BN). Twisting the *h*-BN layer by 180 degrees matters (marked by '), adding three high symmetry stacking configurations, as shown in Fig 1c. Contrary to the unstable AA in Fig 1b, the most abounded and stable BN polytype is the fully eclipsed antiparallel AA' (also called 2H)[45–48,60], indicating that eclipsed atomic sites are favored over hollow sites, unlike in graphene. Due to different electron affinities, the Coulomb attractions of eclipsed and oppositely charged boron\nitrogen atoms pull the layers closer together (while the Coulomb attraction in the hollow AB sites cancels by symmetry[4]). The second potential well in Fig.1c corresponds to a meta-stable AB1' stacking configuration eclipsing two boron (red) atoms, while the nitrogen (blue) eclipsed configuration AB2' is unstable [45]. Here, electron redistribution within each monolayer results in smaller boron sites than nitrogen's, justifying the relative stabilities [61,62]. Similar considerations apply to honeycomb TMDs where one site includes two chalcogen atoms rather than one[63]. The overall stability depends on the depth of the vdW potential wells (fig. 1b.c), the sliding distance between them, and the planar stiffness. Thus, non-honeycomb polytypes with lower symmetry, slightly shifted layer positions ($u_x \ll a$), and shallower vdW wells are less stable [64], and tend to lose the stacking order below room temperature [65,66]. Similarly, changing the lattice constant by $\gtrsim 1\%$ or the relative twist angle between the layers by $\gtrsim 3$ degrees flattens the vdW potential curve and eliminates the structural stability[49,67–69].

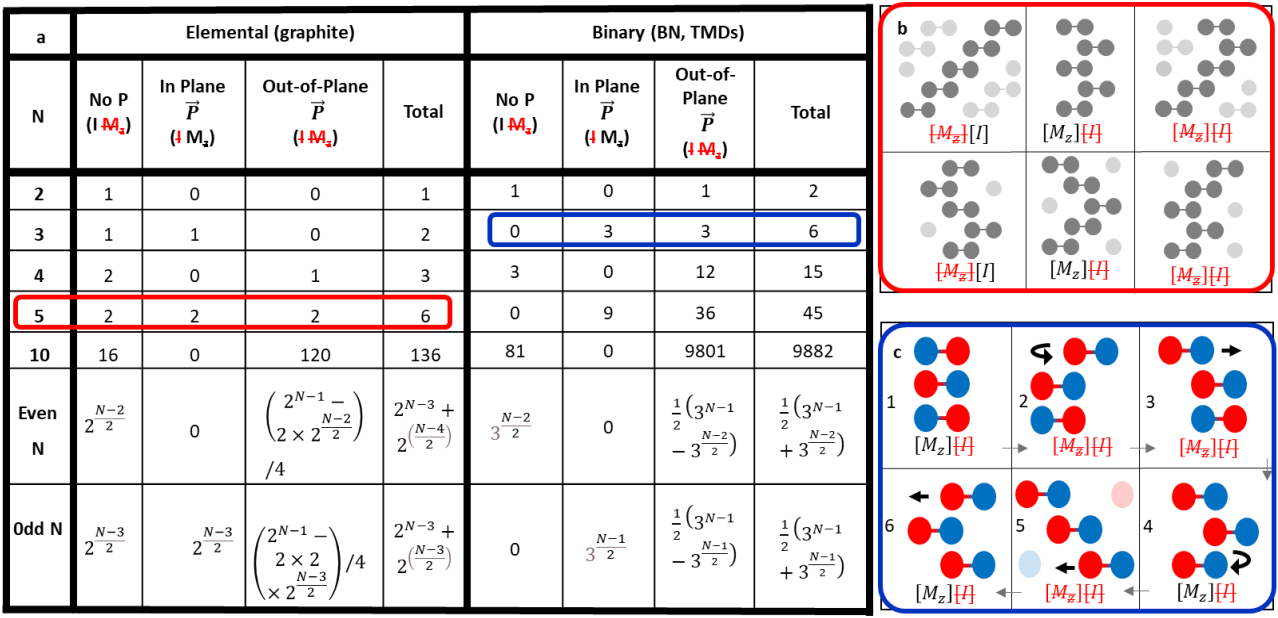

*Figure 2: Sorting polytypes and symmetries.* (a) The number of distinct polytypes in N layers of graphene[22] and honeycomb binary compounds, sorted into their inversion [I] and mirror-plane [$M_z$] symmetries. The count considers metastable AB/BA/AA' interfaces, which are typically stable on macroscopic scales. (b) Side view of the six possible polytypes in penta-layers graphene and (c) in tri-layers of honeycomb binary compounds. Transitions between polytypes 1 to 6 by $180°$ rotation or by $a$-shift translations are marked by arrows.

These primary adhesion considerations extend to multilayer polytypes with additional weakly interacting interfaces. Each extra interface contributes two metastable stacking configurations, named by the cyclic layer positions A, B, and C in Fig.1d [70–72]. Trilayer graphene, for example, forms two distinct periodic crystals that qualify as (meta) stable 2D polytypes: ABA (Bernal, B) and ABC (Rhombohedral, R). While ABA is the prevalent polytype, ABC domains are frequently observed in natural graphitic flakes, indicating the meta-stability of multilayer polytypes that avoid AA stacking[42,73]. The table in Fig. 2a sorts the elemental 2D vdW polytypes with one type of atom (like graphite). Note that many of the $2^{N-1}$ configurations are merely different orientations of equivalent crys-tals[22]. For example, the eight possible stacking configurations of tetra-layers form three distinct polytypes, while penta-layers can slide between six polytypes (Fig.2b). The number of binary-compound polytypes is even larger (see the second column in Fig.2a), with each antiparallel interface adding a stable AA' configuration that typically dominates the AB1'. Thus, binary trilayers host six distinct polytypes (Fig.2c) that cover macroscopic structures.

**Poly-symmetries**
Honeycomb graphene monolayers have sixfold rotation and inversion [I] (x→-x, y→-y, z→-z) symmetry points. Adding a second AB layer reduces the rotation symmetry to threefold, and ABA trilayers further break [I], thus allowing internal electric polarizations (Fig.2a). Binary compound honeycombs break [I] already at the monolayer level but retrieve it in AA' polytypes with even N[74]. Once the system loses an inversion center, the mirror symmetry $M_z$ (z→-z) determines the out-of-plane polarization component $P_z$. For example, the AB/BA polytype of BN[4,5,8,75,76] or TMDs[6,7,10,11] exhibit opposite $P_z$ at room temperature owing to broken [I] and $M_z$. A notable symmetry of binary AB bilayers is that interlayer sliding is equivalent to $M_z$ in switching $P_z$ upside down (see in Fig. 1b how AB switches to BA by either $a$ shift or $M_z$). Thus, applying external electric fields to switch $P_z$ is a direct mechanism to control the in-plane sliding and vice versa. Note the fundamental difference from typical ferroelectric materials in which atomic motions are primarily along the field direction [77–79].

Fig. 2a sorts the polytypes by [I] and $M_z$ symmetries, pointing to allowed internal polarizations. As shown, even-N polytypes that break [I] also break $M_z$ and may host finite $P_z$. The thinnest elemental polytype to do so is ABCB tetralayer graphene, exhibiting $P_z$ at room temperature [22]. While polytypes that break [I] and preserve $M_z$ exhibit a second harmonic optical response [80], all polytypes are uniquely defined by distinct symmetries and dispersions. In graphitic penta-layers, for example (Fig.2b), two non-polar polytypes (out of six) preserve [I], two polar polytypes preserve $M_z$, and two may host out-of-plane polarization.

**Poly-properties**
The discrete translation symmetry enables reciprocal space perspectives of the polytype's properties. Each orbital in the unit cell contributes one electronic band that shifts in energy relative to the monolayer bands depending on the interlayer overlaps and the symmetries. As a pedagogical model, consider the tight-binding description of graphite with well-established orbital overlaps and intra/interlayer hopping parameters [81]. Each monolayer contains two significant orbitals per unit cell overlapping by $\gamma_0 \approx 3$ eV, and two energy bands that cross at the hexagonal Brillouin zone corners (Fig. 3a), at lattice momentums $K, K' = (\frac{2\pi}{3a}, \pm\frac{2\pi}{3\sqrt{3}a})$. Super-positioning a second AB layer split two of the bands by $\pm\gamma_1 \approx 0.3$ eV away from the Fermi energy (see the red bands). In real space, the split band's momentum states reside on eclipsed dimer pairs (dashed red frames), while the states at the Fermi energy reside on hollow monomers [56,72]. Stacking more Rhombohedral layers adds dimer bands at $\pm\gamma_1$ (red bands in Fig 3.b,c), and further separate the surface monomers. The latter separation suppresses the band's dispersion by $E \propto k^N$ with $k$ measured relative to the K point [82,83]. Notably, at the Fermi energy where $k \to 0$, the electronic group velocity $v_F = \frac{\partial E}{\partial k}$ slows down (proportional to $k^{N-1}$), and for N>2, the density of states $DOS \propto 1/k^{N-2}$ diverges (Fig. 3d). This divergence in rhombohedral polytypes indicates relatively flat bands that host correlated electronic phases like orbital magnetism[16,17], unconventional superconductivity[18–20], and fractional anomalous hall states[21].

Unlike rhombohedral polytypes with one pair of monomers (and a pair of bands near the Fermi energy), Bernal polytypes have N monomers, and all other polytypes are intermediate. Fig 3.b shows a tetralayer Bernal with four monomers and a tetramer, aside ABCB polytype with three monomers, a dimer, and a trimer. Consequently, the four bands of the Bernal tetramer split by ~$\pm\gamma_1, \pm 2\gamma_1$, (green curves) while the ABCB trimer bands split by ~$\pm\sqrt{2}\gamma_1, 0$ (blue). These energy splitting govern the optical response in the infra-red (IR) regime [84–86] (see Fig 3e) and the Raman scattering spectrum up to the visible range (Fig. 3f, left), among other properties. A more accurate tight-binding description includes additional hopping between atoms from the same sublattice, like $\gamma_2 \approx 20$meV, and $\gamma_5 \approx$ 40 meV[81] that preserves momentum but breaks the electron-hole symmetry.[87] These hopping amplitudes open energy gaps and are essential for finite out-of-plane polarization at charge neutrality in polytypes that break [I] and $M_z$. For example, $P_z$ in ABCB polytypes was recently measured using surface potential imaging[22]. In Fig. 3f, the left Raman map distinguishes the Rhombohedral, Bernal, and Polar polytypes, while the right-side surface potential map further distinguishes Polar domains with opposite polarizations. The polarization value confirmed the significance of onsite Coulomb interactions and self-consistent tight-binding calculations [88,89]. Owing to the thin semi-metallic graphitic nature, gate electrodes can further tune the band structure near the Fermi energy [42,90,91]. Thus, adding a structural control to switch between the poly-properties provides multiferroic opportunities that retain a permanent, non-volatile response after removing the external fields.

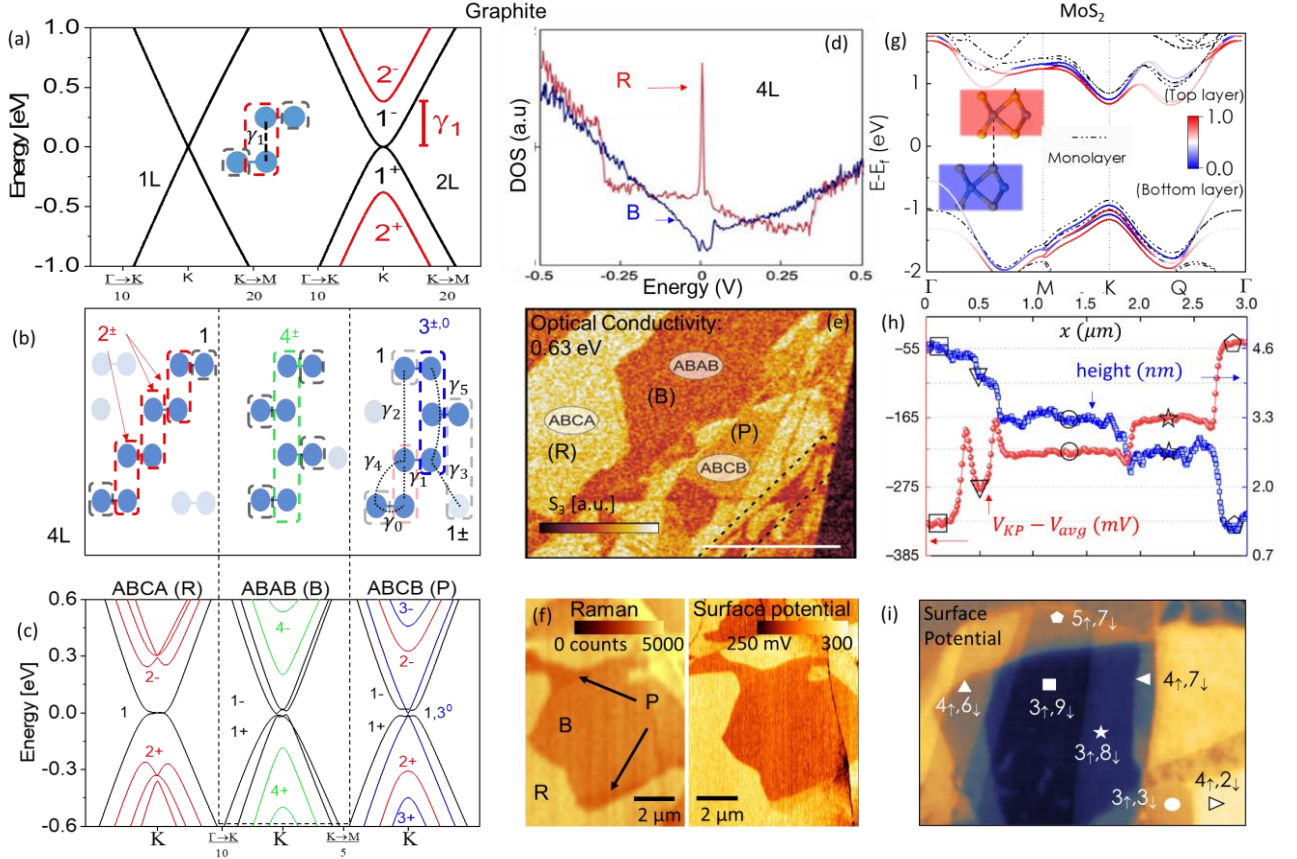

*Figure 3 Poly-properties. (a) Band structure of monolayer/bilayer graphene near the K point at the Brillouin zone edge. In bilayers, touching/split bands' states near K reside on hollow/eclipsed (monomers/dimer) orbitals marked by dashed black/red frames. The energy split relative to the Fermi level, noted by a vertical bar, measures the hopping integral $\gamma_1$. (b) Unit cells of three possible tetra-layer polytypes with relevant $\gamma_{0-5}$ hopping processes. Frames mark mono/di/tri and tetra-mer overlaps of the black/red/blue and green bands near the K point plotted in (c)[22]. Bonding (+)/antibonding (-) states are marked at the valence/conduction bands. R and P polytypes include flat bands over a large momentum range near the Fermi energy. (d) Tunneling current measurements in tetra-layers indicating a diverging density of states (DOS) in the rhombohedral polytype (R)[92]. (e) Scanning near-field optical microscopy at incident photons' energy of 0.63 eV showing distinct scattering intensities in three tetralayer polytypes.[84] (f) The three polytypes are also distinguished by integrated Raman Intensity (left), while the surface potential map (right) further distinguishes the oppositely polarized orientations of the polar polytype (marked in arrows)[22]. (g) Band structures calculations of $MoS_2$ monolayer (dashed black) and parallel bilayers (colored curves). Red/blue colors represent the weight of each momentum state on the top/bottom layers according to the unit-cell configuration illustrated at the inset. While states from the valence band edge near Γ momentum are white (spread on both layers), states near K momentum reside on a particular layer[10]. (h) Surface potential (blue, right axes) and topography (red, left axes) steps in multilayer rhombohedral $MoS_2$ polytypes. The horizontal greed emphasizes that with each added layer and interface, a fixed internal potential step accumulates in a "ladder-like" manner[10]. (i) Surface potential map of mixed stacking configurations in parallel $MoS_2$ polytypes. The surface potential value at each region reveals the number of up-pointing minus down-pointing ($N_\uparrow - N_\downarrow$) interfacial polarizations[10].*

Binary compound polytypes like BN and TMDs host in-plane piezoelectricity at the monolayer level[93,94], direct band gaps[95] (dashed black in Fig. 3g), and pronounced second-harmonic optical response[96,97]. Super-positioning a second antiparallel layer in these polytypes retrieves a centrosymmetric unit cell, which nullifies the polar properties for all even-N structures[74]. Conversely, parallel binary multilayers remain non-centrosymmetric and exhibit out-of-plane $P_z$ in case of broken $M_z$[4–6,10,11,76,98]. Here, inequivalent orbital overlaps promote interlayer charge transfer of $\sim 2\times10^{12}$ $e$/cm$^2$ that shift the potential between the layers by $\Delta V \sim 100$ $mV$[75,99]. In reciprocal space, the monolayers' bands split into bonding\antibonding bands at the center of the Brillouin zone (see the nearly transparent bands in Fig.3g at Γ momentum) that spread over the two layers in real space[10]. Conversely, near the K momentum, each state resides separately on the top layer (red bands) or the bottom layers (blue bands), while the relative bands' energies shift by $\Delta V$.

Notably, in multilayer polytypes, $\Delta V$ was found to accumulate with the number of Rhombohedral interfaces, as shown in (Fig.3h,i). The fixed potential steps, corresponding to the number of up-pointing minus the down-pointing ($N_\uparrow - N_\downarrow$) interfacial polarizations confirm that these "ladder ferroelectrics" evade depolarization effects and hence are free from surface reconstructions, atomic mixing, or external adsorbents, which are unavoidable in thin 3D ferroelectrics. Eventually, the potential difference in thick Rhombohedral flakes saturates once it approaches the bandgap value[100]. Thus, this purely electronic response to the so-called "polar catastrophe" indicates a formation of electron surface-states residing on one surface while valence hole states accumulate on the opposite surface, and add appealing functionalities to the discrete ladder-like response.

In addition to band splitting and internal charge distributions, we emphasize that the structural order couples to many degrees of freedom. For example, optical excitons [101–103], ferro/antiferromagnetic phases[12–14,104], orbital magnetism[17,105], and superconductivity [18,106], were recently reported in honeycomb polytypes depending on the stacking configuration, while other predictions include quantum geometric properties[107–109]. The term "Slidetronics" was introduced [4] to highlight these multiferroic opportunities in which an external electric field controls the structural transitions and subsequently switches a supplemental order.

**Poly-Stabilities**
To discuss the structural stability, switching dynamics, and transition temperatures $T_c$ to a turbostratic structure in which the layers slip out of alignment, we refer to the adhesion potential curves in Fig. 1b,c. The barrier height $E_b$ at the intermediate saddle points (SP) is a couple of meV per atom [42–44], separating two minima that defer by merely $\Delta E_p \sim 1$meV per atom [45,46]. These relatively small energy differences explain the common appearance of adjacent polytypes in naturally exfoliated or thermodynamically grown flakes [10,47,84,110–112] (see Fig. 3e,f,i), and their typical switching dynamics. The favored stacking energy is evident from the prevalence of each polytype. In graphitic tetralayers, for example, the Bernal, Rhombohedral, and ABCB polytypes occupy ~80%, 20%, <1% of the exfoliated flakes, respectively, and remain stable at ambient conditions[22,84,86]. Notably, it is possible to tune $\Delta E_p$ to favor one polytype over the other. For example, pressure tends to favor the Bernal polytypes, while electrostatic hole-doping may favor rhombohedral-like polytypes [113,114].

The turbostratic $T_c$ further depends on the shift displacement between the wells, $\delta u_x$ ($a$ in honeycombs), and the in-plane rigidity of the monolayers. At room temperature (where $K_B T \sim 30$ meV) thermal fluctuations exceed $\Delta E_p$ and $E_b$ per atom. Nevertheless, the fluctuations cannot pull individual atoms out of the wells owing to the monolayer's stiffness with shear $E_G$ and Young $E_Y$ modulus ranging from 100GPa to 1Tpa[115–117]. Rather, the interfacial misalignment stretches over partial dislocation lines of width W [118,119]. Consequently, the transition energy scale, $K_B T_c \sim E_b \left(\frac{W}{a}\right)^2$, is set by collectively incommensurate atoms in the boundary dislocation strip, separating perfectly commensurate strain-less and twist-less domains (see Figure 4a). On the one hand, narrow $W$ reduces the number of incommensurate atoms across the dislocation, losing a collective adhesion energy $\sim E_b W/a$. On the other hand, the strain energy of narrow dislocation strips grows by $\propto \delta u_x^2/W$.

The compromise gives $W \approx \frac{\delta u_x}{2}\sqrt{\frac{1}{E_b}\left(\frac{E_Y}{1-\nu^2}\sin^2\varphi + E_G\cos^2\varphi\right)} \approx 30\delta u_x$ with $\nu$ the Poisson ratio and $\varphi$ the angle of the dislocation strip measured from the armchair direction [44,118]. Typical strip energy cost is ~ 1 eV per nm length in graphite[118], BN [120], and TMDs [119,121–123], with shear dislocations running along armchair favored over tensile/compressive dislocations along zigzag. Thus, honeycomb polytype with $\delta u_x = a$ are stable up to several eV, much above room temperature[124,125]. Conversely, non-honeycomb stacking polytypes like the distorted 1T phase of TaS$_2$ exhibiting charge density waves or the polar T$_d$ polymorph of WTe$_2$ have $\delta u_x \ll a$ dislocation vectors, narrower $W$ ~ 1nm, [64] and significantly lower T$_c$[65,66,125]. We note that a detailed analysis of this 2D phase transition (of the Berezinskii Kosterlitz-Thouless type[126,127]) should include the full anisotropic stacking potential and a vortex/antivortex dislocation's formation as in thin oxide films[128].

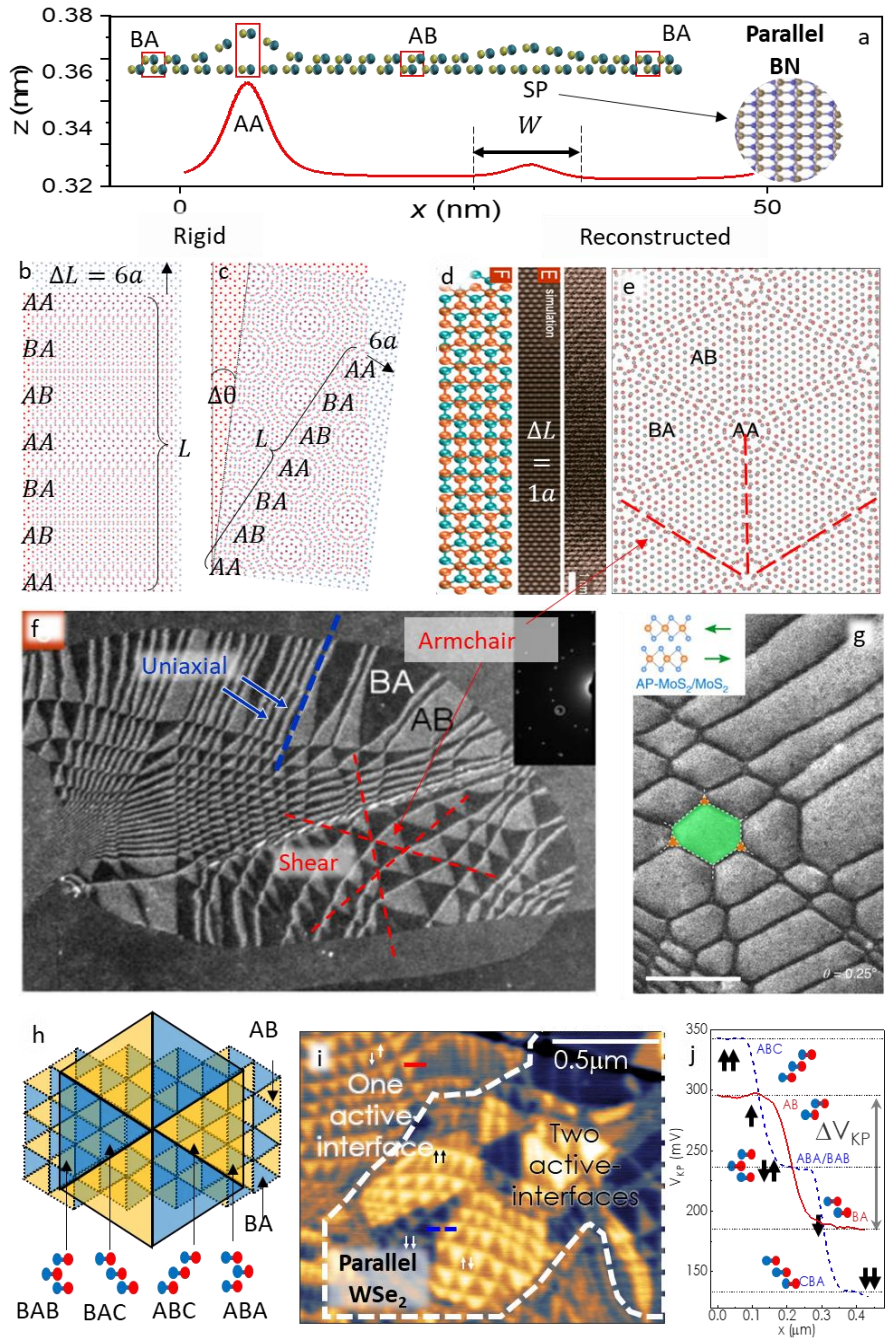

*Figure 4: Adjacent polytypes by artificial stacking. (a) Line-cut (red) of the calculated interlayer separation Z along partial dislocations in a parallel BN bilayer. Z grows by ~ 10 % at AA dislocations and 1% at SP strips (right inset)* [4]. *Top inset: exaggerated cross-section illustration. (b/c) A relative uniaxial strain/twist ($\Delta\theta$), between rigid honeycomb bilayers producing a 6a interlayer shift per L. (d) Top view simulation and TEM imaging of a uniaxial partial dislocation ($u_x = a$) in bilayer graphene.*[118] *(e) Top view illustration of partial shear dislocations in bilayer graphene*[132]. *(f) Dark-field TEM image of marginally-twisted bilayer graphene* [118]. *Uniaxial/shear dislocations identified according to the lattice orientations of the dislocation strips are marked in blue/red (g) Dark field SEM imaging of antiparallel bilayer MoS$_2$. AA'/AB1' regions are marked in green/orange* [123]. *(h,i,j) Illustration, surface potential map, and typical surface potential line cuts of parallel bilayer (red) and trilayer (dashed blue) polytypes of WSe$_2$* [10]. *The cumulative potentials and moiré landscape identify all polytypes distinctively.*

**Poly-assembly**

The metastability of commensurate honeycomb interfaces guarantees a reliable assembly of all their possible polytype configurations in case of marginal interlayer twist, relative strain, or lattice mismatch. To see how, consider the overlaps of two rigid layers shown in Fig. 4b after stretching one by $\Delta L = 6a$ along the armchair direction. This rigid stretch creates a 1D moiré pattern with two cycles of $L/6$ - wide incommensurate AA, AB, BA – like configurations.

In practice, however, for small strains, such that $\Delta L/L < a/W$, with L the structure's dimension, this rigid moiré picture breaks. Rather, the deformation is confined within a narrow boundary strip (Fig. 4d), separating two fully commensurate unstrained polytype regions. Similarly, Fig. 4c shows a rigid twist with interlayer displacement $\Delta L = \theta L$ setting a critical deformation angle $\theta_c < \frac{a}{W} \sim 1°$. Once the moiré supercell periodicity exceeds $W$, the stable AB and BA regions grow into triangular domains while the incommensurate AA-like regions shrink into local ~ $W^2$ points (Fig.4e)[118]. Since AB, and BA are mirror transformations of the same polytype with $\Delta E_p$=0, the triangles cover similar areas as shown in Fig. 4f. Conversely, in antiparallel binary compounds, the most stable AA' polytype (see Fig. 1c) grows into larger regions (see the green area in Fig. 4g), the AB1' domains remain intermediate (orange), and the unstable AB2' regions shrink into local dislocation points [123]. Hence, marginally twisted stacking not only imposes adjacent domains of all possible polytype configurations (per parallel/antiparallel family) but also distinguishes their domain area and shapes.[113,129] In particular, the relative stability of each polytype as a function of temperature, electric fields, pressure, or strain can be extracted by analyzing the relative area of the domains and the curvatures of the dislocation strips.[23,113,129–131]

Polar binary polytypes are further recognized from their cumulative surface potentials. For example, marginal twisted TMD trilayers guarantee four configurations of three distinct polytypes ↑↓, ↓↑ and ↑↑ (or ↓↓) as marked by arrows in Fig. 4h, corresponding to the interfacial polarizations and surface potentials shown in Fig. 4i,j.[10] We note that in addition to mechanical stamping, the monolayer nucleation conditions during the growth may impose the polytype orientation, for example, using surface terraces of the substrate [133–136]. More generally, poly-stability and decisive assembly opportunities also apply to non-identical layers if the lattice mismatch is below ~1% [137,138] and the twist angle is negligible. Graphen/hBN interfaces (1.5% mismatch), for example, are borderline with weakly commensurate honeycomb moiré patterns (~ $100a$ periodicity). In this case, finite metastability was observed for $\theta_c < 0.1°$ [69].

**Poly Switching**

While the energy barrier associated with breaking covalent bonds limits practical switching between graphite and diamond, [27–30] structural switching into amorphous phases of alloys is considered applicable [139]. In response to local electric and optical pulses, phase-change 3D alloys heat up above their melting point and cool down slowly /rapidly to determine a recrystallized/amorphous phase [140]. Another type of phase-change material is distorted TMDs, in which tiny bond shifts within each layer reduce the honeycomb symmetry in response to strain [39] or heavy doping [40]. Although enabling purely electrical switching between layered polymorphs (thinner than the field penetration depth), the relatively large switching fields reported to date pose practical challenges. In this sense, sliding undistorted layers against relatively small energy barriers of $E_b \sim 1$meV per atom in polytypes is promising for efficient, reversible, and rapid electric switching between multiple polytype configurations at room temperature despite much higher crystallization temperatures.[124,125,141,142]

The polytypes switching mechanism involves sliding boundary dislocation strips in a soliton-like motion across the system, leaving the new structural configuration behind. For example, Fig. 5a shows a local shear stress applied from an AFM tip, pushing a boundary dislocation to extend an ABC graphene polytype over ABA region [143]. Dislocation motion and sliding ferroelectric switching in response to external electric fields were first inferred from hysteretic conductivity curves of distorted WTe$_2$ tri/bilayers (Fig 5b) [3]. The electrically driven sliding of boundary strips (rather than a rigid layer motion) was reported in polar BN polytypes by imaging the electric surface potential, enabling detailed information on the switching dynamics [4] (Fig 5c). As noted, by avoiding pining sources such as dense moiré networks, disorder, and edge open bonds [23], the partial dislocation strips show high mobility, with coercive fields as low as ~ 0.1V/nm [4–7], and sliding characteristics that resemble the superlubricant motion of incommensurate vdW interfaces[49,68]. Notably, electric-field control of the stability and switching dynamics extends to non-polar polytypes[113,129]. Fig 5d shows $M_z$ symmetric ABA and ABC moiré domains of graphite, which expend/shrink in response to external fields, although $P_z$ =0 in both polytypes. Here, uneven doping of the top and bottom layers due to the electric displacement field is sufficient to modify the sign of $\Delta E_p$, deform the boundary strip network, and extend the more stable polytype configuration. We further note that in dense moiré networks, the response is not hysteretic since the network's tension pulls back the elongated strips once the external electric field is removed[23]. Therefore, substantial twists result in a paraelectric-like response, while the hysteretic response is reserved for minute twist angles in which the boundary strip's motion becomes independent, as shown in Fig 5e.

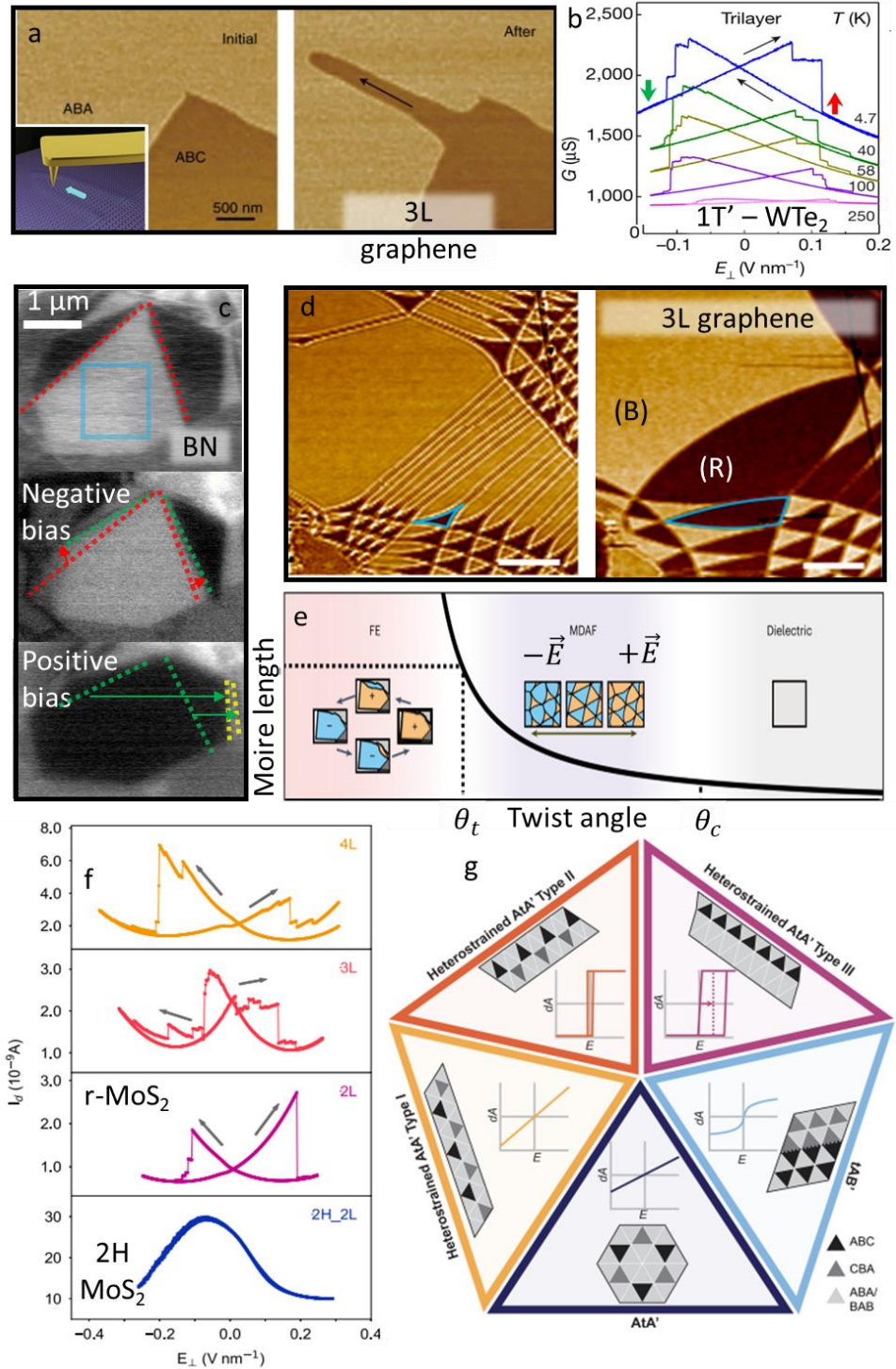

*Fig 5: Poly-switching dynamics by sliding partial dislocations strips.* (a) Deforming ABA/ABC boundary dislocations in trilayer graphene using mechanical shear force from an AFM tip[143] (b) Hysteretic curves of conductance versus electric displacement field in distorted 1T' WTe$_2$ [3]. (c) Hysteretic electric switching of polar BN domains by a biased AFM tip [4]. (d) Non-hysteretic electric deformation of the moiré dislocation network between non-polar graphitic polytypes captured by near-field optical microscopy [144] (e) Hysteretic to non-hysteretic transition in dense dislocation networks at a characteristic twist angle $\theta_t$.[23] (f) Hysteretic conductance curves indicating electric switching of polar bi/tri/tetralayer MoS$_2$ polytypes [11]. (g) Polystructural switching and electric susceptibilities by interacting dislocation strips in polar trilayers. [24]

In multilayer polytypes, the motion of dislocations from separate interfaces leads to switching between more than two metastable configurations. In ladder-ferroelectrics[10], for example, electric switching between several polarization states was inferred from current hysteretic curves appearing up to tetralayer structures with three polar interfaces [11] (Fig. 5f). Moreover, dislocation strips from successive interfaces may interact elastically to pull/repel each other. For example, the dislocations from adjacent interfaces tend to align in consecutive twists. In contrast, strips in oppositely-twisted successive interfaces (where the first and third layers are nearly parallel) may repel [24]. Consequently, the overall electric susceptibility, coercive fields, and switching dynamics show a wide range of responses, as shown in Fig 5g, which are unavailable in 3D ferroelectrics.

**Outlook, Bottlenecks, and Poly-Opportunities**
Looking ahead, we expect further studies in a broad range of layered compounds, focusing on the impact of polytype configuration on collective crystal excitation. How can one improve the interlayer coupling? Boost the electronic band hybridizations near the Fermi energy? Better couple the spin and orbital orders to the structural symmetry? Model planar and vertical charge redistributions? are vital questions for ongoing theoretical analysis, modeling, and experiments. The intuition developed in elemental honeycomb and binary compounds should carry on to more complex monolayer building blocks with additional compounds and reduced symmetries. We note the vast space of structural coupling opportunities in "SlideTronics" that adds up to the switching of electric conductivity ("ElecTronics"), spin-dependent transport ("SpinTronics") [145], and valley occupations ("ValleyTronics") [146]. The first step is to study low-hanging fruits by comparing newly identified meta (stable) polytypes. Note that metastability and in-plane relaxations in mechanically assembled layers guarantee detailed comparisons of all possible polytypes and structural orientations in adjacent domains, which are insensitive to environmental perturbations. We expect additional recipes reporting ways to identify the adjacent polytypes using optical / Raman imaging/ dark-field electron microscopy / electric surface potential or magnetic imaging, preferably at room temperature. More complex devices will follow once the most rewarding polytypes and responsive structural transitions are recognized.

The periodic symmetry maintains all the information within a single unit cell, which dramatically simplifies the above analysis. However, the coupled electronic and elastic response at the boundary strips is a great challenge that calls for intermediate-scale approximations, including a few hundred atoms. Experimental-wise, the interactions between strips within one interface or from adjacent interfaces shade light on the local properties and can be detected by tracking the strip's dynamics. Similarly, local information on the density of states and atomic positions across the boundary dislocation is vital to amend the models' parameters.

A central bottleneck for applications is barriers to nucleate domains on-demand using local stimuli rather than pushing preexisting strips, as reported so far. Since the polytype stability and domain nucleation energy depends on the system's dimensions, shrinking the devices towards the $W$ scale should reduce the associated coercive fields. Conversely, cutting the structure into local single-domain islands introduces edge-open bonds that zip the layers together and pose a pinning barrier. This conflict might be addressed by passivating the edges or bypassing their impact. A prominent device concept that bypasses this conflict is currently perused by sagging the functional layers into local cavities that form active polytype islands (without cutting the functional layers)[26]. Markedly, the cavities can be etched into super lubricant spacers that support long-range elastic interactions between the crystalline islands. Compared to 2D walls in 3D phase-change materials, the 1D nature of the boundary strips provides outstanding flexibility to control the intermediate domain pattern and switching dynamics [147,148]. Designing the planar device dimensions and shape can stabilize intermediate transition patterns, topologies (number of strips that cross the device), and switching coercive fields. Electrical readout of the 2D device state is enabled by resonant tunneling schemes across ultimately two-atom-thin internal polarizations [149], and in the plane as in ferroelectric field effect transistors[3,25].

Beyond structural optimization, it is intriguing to combine external perturbations for switching. Mixing local light pulses, charge currents, stress, and displacement fields may optimize the strip nuclea-tion and motion for a given application. We note that major durability challenges in structural transitions of 3D bonded crystals due to atomic reconstruction and irreversible chemistry at the surface dangling bonds are naturally avoided in layered surfaces [141,142]. Finally, scaling up and cost-effective competition with well-established silicon-based technologies are crucial. While field-effect transistors and numerous vdW heterostructures[150] based on TMDs have already been introduced as part of a practical technology roadmap [151], polytype-based devices require the growth or assembly of uniform lattice orientation, preferably on wafer scales [152–154]. Slide vdW polytypes confront these massive scaling-up challenges with visionary state-of-the-art functionalities, including non-volatile multi-ferroelectric tunnel junctions, cumulative polarizations, and elastically coupled domains that enable numerous novel device concepts.